\documentstyle[12pt,a4,/user1/scol/bin/new]{article}
\begin{document}

\bibliographystyle{unsrt}

\begin{center}

{\Large\bf {On O($N$)-Symmetric Gauged $\phi^6{}_{2+1}$ Theory
with Chern-Simons Term}}

\vspace{2cm}

S. H. Park

Research Department\\
Electronics and Telecommunications Research Institute\\
P.O. Box 106, Yusong-gu, Taejon, 305-350, Korea\\
E-mail: shpark@logos.etri.re.kr

\vspace{2cm}

\bf{Abstract}

\end{center}

I investigate the effects of the Chern-Simons coupling on high-energy
behavior in $2+1$ dimensional U(1) gauged
$\eta (\phi^\dagger \phi)^3$ theory
with a Chern-Simons term. The effective potential and the
$\beta$ function for $\eta$ are calculated to the next-to-leading order
of the $1/N$ expansion as functions of $\theta$ (the Chern-Simons coupling).
For all $\theta$, the theory
is found to be driven to instability region at high momenta.
It is briefly discussed on the radiative corrections to $\theta$.
\pagebreak

It has been suggested \cite{bmb} that $\eta \phi^6{}_{2+1}$ theory reveals
interesting nonperturbative phenomena. The authors in \cite{bmb}
argued that the theory shows a non-trivial ultraviolet (UV) fixed point at
$\eta = \eta_c \equiv (4 \pi)^2$, where the mass is dynamically
generated so that the scale symmetry of the
theory is spontaneously broken,
in a nonperturbative method.
Their new phase arises at the leading order in the $1/N$
expansion where the perturbative $\beta$ function vanishes.
The high energy behavior of the
theory is governed by $\eta_c$. The UV fixed point at
$\eta = 192 > \eta_c$
previously found at the next-to-leading order of the $1/N$
expansion \cite{pis} lies in the unstable region, which is
dominated by their stable new phase.

Regarding this, however, several analysis have indicated that
this new phase indeed shows
instability at a finite $N$: at large momenta $\eta$ is driven to the
instability region by its renormalization group equation at the
next-to-leading
order in the $1/N$ expansion \cite{gudm}.
A numerical analysis also shows the
instability of $\eta_c$ at finite $N$ \cite{kars}.

In this paper I investigate the same phenomena in the U(1) gauged
$\eta (\phi^\dagger\phi)^3{}_{2+1}$ theory, where $\phi$ is a complex
field. The dynamics of the gauge field is given by the Chern-Simons (CS)
term. The theory (Chern-Simons-Higgs model) possesses a new soliton
solution for a completely symmetric $\phi^6$ potential which is
distinguished from the usual one in the Maxwell theory with $\phi^4$
potential \cite{kim}. With the CS term the gauge sector of the theory is
renormalizable (though not superrenormalizable)
in $2+1$ dimensions, which makes it
field theoretically interesting. While the CS term has been known to
affect the long-distance behavior of the theory, the motivation in
this paper is to study the CS effects on the short distance behavior
\cite{park}

An explicit $\theta$ dependence of the effective potential and the
renormalization group flow for $\eta$ will be
given at the next-to-leading order of $1/N$ expansion.
The $\beta$ function for $\eta$ depends on $\theta$.
However, it does not qualitatively change
the original picture of the $\eta \phi^6$ coupling in \cite{bmb}:
$\eta$ is driven to
instability
region at high momenta for any value of $\theta$.
The renormalization of $\theta$ [7,8]
will be also shortly mentioned.

I consider the gauged $(\phi^\dagger\phi)^3$ model with a Chern-Simons
term in $2+1$ dimensions which is given by (in euclidean version)
\begin{equation}
{\cal L} = \overline{D_\mu\phi^i} D^\mu\phi^i +  \phi^{i\dagger}
\phi^i + {\lambda \over 4} (\phi^{i\dagger}\phi^i)^2 + {\eta \over 3}
(\phi^{i\dagger}\phi^i)^3 + i \theta \epsilon_{\mu\rho\nu}A_\mu \partial_\rho
A_\nu,
\end{equation}
where $i=1,..,N$ and $D_\mu = \partial_\mu + iA_\mu$ \cite{statement}.

Intruducing two auxiliary fields $\chi$ and $\sigma$ and
rescaling the couplings $m^2, \lambda$, $\eta$, and $\theta$ to facilitate
the $1/N$ expansion eq.(1) is rewritten as
\begin{equation}
{\cal L} =  \overline{D_\mu\phi^i} D^\mu\phi^i + \sigma(\phi^{i\dagger}\phi^i -
N \chi) + N (m^2 \chi + {\lambda \over 2} \chi^2 + {\eta \over 3} \chi^3)
+ iN\epsilon_{\mu\rho\nu}A_\mu \partial_\rho A_\nu.
\end{equation}

The effective potential is obtained by Legendre
transformations of the energy density with respect to the external
sources introduced in the Green's function generator. The detailed
derivation is referred to [10,3].
For $<A_\mu> = 0$, there is no gauge field contribution
at the leading order, therefore
the effective potential at $O(N)$ is then the summation of the $\phi$-loop
diagrams with the classical parts,
\begin{equation}
V^0 (\vert \phi \vert, \chi, \sigma)~ =~ N \int {d^3p \over (2 \pi)^3}
{\rm ln} (p^2 + \sigma) +  N (m^2 \chi + {\lambda \over 2} \chi^2 + {\eta \over
3} \chi^3)
+  \sigma( \vert \phi \vert^2 - N \chi).
\end{equation}

To derive the $1/N$ expansion in the eq (2) one has to
consider the vacuum structure which is characterized by the following gap
equation
\begin{equation}
{\partial V^0 \over \partial \sigma}(\vert \phi \vert, \chi, \sigma) =
N  \int {d^3p \over (2 \pi)^3} {1 \over p^2 + \sigma} + \vert \phi \vert^2
- N \chi = 0.
\end{equation}

Now it is sufficient to choose the zero expectation value of the $\phi$
field, the symmetric phase, to study the high energy behavior of the
theory. I refer to \cite{gudm} for the detailed phase
structure of the theory. Eliminating the unphysical field $\sigma$
using the above gap equation one obtains the following effective
potential in the symmetric phase
\begin{equation}
V^0 (\chi) ~=~ N [{1 \over 3} ({\eta - (4 \pi)^2}) \chi^3 + {1 \over 2}
\lambda \chi^2 + m^2 \chi].
\end{equation}
In the above equation, the parameters $\chi, \lambda$, and $m^2$ are
renormalized ones. In particular, the renormalized $\chi$ is obtained
through
\begin{equation}
\chi = \chi_0 - {\Lambda \over 2 \pi^2},
\end{equation}
where $\chi_0$ is the bare parameter and $\Lambda$ is the cutoff of
the theory.

In terms of the renormalized parameters the gap equation,
eq. (4), is (in the symmetric phase)
\begin{equation}
-\chi - {{\sqrt \sigma} \over 4 \pi} ~=~ 0
\end{equation}
Therefore, it is required $\chi \leq 0$.
Note that $\eta$ remains finite at the leading
order of $1/N$.
Now the stability condition at the leading order of the $1/N$ expansion,
 $0 \leq \eta \leq (4 \pi)^2$, is reobtained from the effective
 potential.

The photon contributes to the higer order corrections in $1/N$.
The resummation technique of the $1/N$ expansion results in
the inverse photon propagator
\begin{equation}
\Gamma_{\mu \nu} (p^2) ~=~ \Gamma(p^2) (\delta_{\mu \nu} - {p_\mu p_\nu \over
p^2}) - \theta \epsilon_{\mu\rho\nu}p_\rho,
\end{equation}
where
\begin{equation}
\Gamma(p^2) ~=~ {1 \over 2} (p^2 + 4 \sigma){{\rm arctan} {{\sqrt {p^2}}
\over 2 {\sqrt \sigma}} \over 4 \pi {\sqrt {p^2}}} - {{\sqrt \sigma}
\over 4 \pi}.
\end{equation}
The factor $N$ in front of the CS term of eq.(2) has been
absorbed through rescaling of vertices: Vertices are multiplied
by $1/{\sqrt N}$ or by $1/N$ for consistent expansion.

{}From eqs (8,9) one can read that the photon behaves like
$(1 + \theta^2)^{-1} p^{-1}$ (the symmetric part) and
$[\theta/(1+\theta^2)]p^{-1}$ (the asymmetric part) as $p \rightarrow
\infty$. Now the renormalizability of the theory can be proved by
power counting and the Ward identity.
The superficial degree of divergence D of diagrams
is the one from the $\eta \phi^6$ theory without the gauge
field \cite{gudm} plus the gauge field contribution, which is given by
\begin{equation}
D = 3 - {1 \over 2} E_\phi
- E_\chi - 2 E_\sigma - E_{A_\mu},
\end{equation}
where $E_\phi, E_\chi, E_\sigma$, and $E_{A_\mu}$ are the number
of external legs of $\phi, \chi, \sigma$, and $A_\mu$ fields respectively.
One should note that the symmetric part of the photon diagrams is
finite, since the Maxwell term does not appear as a counterterm by
eq. (10), which manifests the renormalizability of the
theory.

By power counting, the CS term is a marginal (not super-renormalizable)
operator, therefore, $\theta$ is a dimensionless coupling which
corresponds to the dimensionless statistics parameter. It seems that
the general argument on the non-renormalization of the
$2+1$ dimensional
topological mass term \cite{coleman}
can be applied to the current model. It is unlikely
to violate the analiticity and the gauge invariance of the theory
which are the main ingredients in the argument of ref. \cite{coleman}
\cite{renormalization}.
It does not, however, mean to rule out the possibility of the finite
radiative corrections to $\theta$ in the asymmetric phase \cite{hlp},
which may be relevant to the
study on the tricritical phenomena of the theory.

The effective potential at the next-to-leading order has
new divergences, and is given by
\begin{equation}
V^1 (\chi, \sigma) = \int \!{d^3p \over (2 \pi)^3}
{\rm ln} \left[ 1 + 2A
\int \!{d^3q \over (2 \pi)^3} {1 \over (p+q)^2 + \sigma}
{1 \over q^2 + \sigma} \right]
+ {1 \over 2} Tr {\rm ln} \Gamma_{\mu \nu},
\end{equation}
where $A = {\lambda \over 2} + \eta \chi$.

Explicitly the last term in the above equation is given by
\begin{equation}
{1 \over 2} Tr {\rm ln} \Gamma_{\mu\nu} ~=~ {1 \over 2} \int \! {d^3p
\over (2 \pi)^3} {\rm ln} \left[\Gamma^2 + \theta^2 p^2\right].
\end{equation}
Expanding the logs in the above equation the UV divergences
can be read off
and then renormalization can be performed introducing
counter term for each divergence.

To get the renormalization group flow for $\eta$
only the logarithemic divergent terms
will be
collected. Now two parameters, $M$ and $\mu$, will be introduced:
$M$ is the mass scale which provides the lower
limit of the divergent integrals in eq. (11) and $\mu$ is the
renormalization parameter on which the parameters
are dependent. As in ref \cite{gudm}
$M$ may be given as
\begin{equation}
M^2 = {\rm max} (\sigma,A^2)
\end{equation}

Collecting the infinities in the eq. (11) and using the gap equation
(7), one obtains
\begin{eqnarray}
V (\chi) &=& V^0 (\chi) + V^1 (\chi) \nonumber\\&&
= \mbox{} N \left[{1 \over 3} (\eta - (4 \pi)^2) \chi^3 + {1 \over 2}
\lambda \chi^2 + m^2 \chi\right] \nonumber\\ && \mbox{}
+ \left[{A^2 \over 2 \pi^2} ( \chi - {A \over
192} )
-{16 \over 3} {1 \over \kappa^2} + ({1 \over 32} + {1 \over 4
\pi^2}) {1 \over \kappa^4} - {1 \over 3(16\pi)^2} {1 \over \kappa^6}
\right]\chi^3 ~{\rm ln} {M \over \mu},
\end{eqnarray}
where $\kappa^2 = \theta^2 + {1 \over 16^2}$.
The appearance of the parameter $\mu$ may be clarified now.
What has happened is that the leading order parameters $\eta, \lambda$,
and $m^2$ are no longer constant values but
should be thought of as functions of $\mu$. The $\mu$ dependences are
all absorbed in the parameters while the effective potential
remains $\mu$ independent.
This argument yields the following
$\beta$ function for $\eta$
\begin{eqnarray}
\beta(\eta) ~=~ {1 \over N} \left[{3 \eta^2 \over 2 \pi^2}
(1 - {\eta \over 192}) - {16 \over \kappa^2} + ({3 \over 32} + {3 \over
4 \pi^2}) {1 \over \kappa^4} - {1 \over (16 \pi)^2}{1 \over \kappa^6}
\right].
\end{eqnarray}
The renormalization group flows of other parameters, $\lambda$
and $m^2$, are not affected by the gauge field so that remain
as in ref. \cite{gudm}.

The $\kappa$ dependent terms are the gauge field contributions.
In the $\kappa \rightarrow \infty$ limit, the gauge degrees of freedom
drop out and the theory is characterized only by the scalar
interactions. This can be readily checked in eqns (14,15).
Even when the dynamics of the gauge field is not given, $\theta = 0$, the
$1/N$ expansion renders the gauge field contribution to the high
energy behavior of the theory.
When $\theta = 0$, the UV fixed point, the biggest
solution of equation $\beta (\eta) = 0$, is larger than 192,
therefore lies in the instability region.

A numerical calculation shows that the UV fixed point
takes the minimun value 174 which is larger than $(4 \pi)^2$.
Therefore, for all $\theta$ the theory is driven to the instability
region
where the energy density is not bounded below
at high momenta. The theory is consistent for any $\theta$
with the $\eta \phi^6$
theory without the CS term.

With the Maxwell term, $F_{\mu\nu}^2$, the photon would not
affect the high energy behavior of the theory.
The photon propagator has ${1 \over p^2}$ behavior instead of
$1 \over p$ at high momenta, which makes the gauge sector
super-renormalizable. It can be easily checked that there
is no $\theta$ dependent log divergences in eq. (11)
\cite{statement2}.

It would be very interesting to investigate the tricritical
phenomena of the theory. It is clear that the CS term severely
affects the scaling behavior at the tricritical point.
It may be possible then to predict the role of the CS term
in the tricritical behavior of condensed matter systems \cite{preparation}.

In summary, I find that the CS term affects the high energy behavior
of the $2+1$ dimensional U(1) gauged $\eta \phi^6$ theory with
a CS term. The nature of the critical coupling found by BMB is
investigated at the next-to-leading order of the $1/N$ expansion.
The effective potential and the $\beta$ function for $\eta$ are
calculated to O(1) and O($1/N$) respectively, which show the
strong dependence of $\theta$. The theory is driven to a
ultravilet fixed point (a function of $\theta$) which is located
at the instability region at high momenta. Therefore, as in the
$\eta \phi^6$ theory without gauge field, the cutoff limit
of the theory has no stable ground state.

\vspace{2cm}

\begin{center}

\underline{\makebox[8cm]{\ }}

Acknowledgement

\end{center}

This work has been initiated with the support from postdoc program of KOSEF.
I would like to appreciate professor J. H. Yee for the hospitality during
my stay in Yonsei university as a postdoc. I am also grateful to the director
of research department at ETRI, Dr. E. H. Lee, for his cooperation on
this research.


\begin{thebibliography}{99}
\bibitem{statement} One may introduce
another igauge coupling $e$ as $D_\mu = \partial_\mu + i e A_\mu$.
However, After rescaling
the $A_\mu$ field one is left with one gauge coupling. Here it is denoted
as $\theta$.
\bibitem{renormalization} In ref. \cite{park1}
the $\theta$ term renormalization of the $2+1$ dimensional
${\rm CP}^{N-1}$ with a $\theta$ term
results in the cancellation of diagrams, which leads to the non-vanishing
$\theta$ flow.
\end{thebibliography}
\end{document}